\documentclass[a4paper]{jpconf}
\usepackage{graphicx}
\begin{document}
\title{First results on the measurements of the proton beam polarization
at internal target  at Nuclotron\footnote{Dedicated to the memory of Prof. L.S.Zolin}}

\author{V~P~Ladygin$^{1\dagger}$, Yu~V~Gurchin$^{1}$, A~Yu~Isupov$^{1}$, M~Janek$^{2}$,  
A~N~Khrenov$^{1}$, P~K~Kurilkin$^{1}$, A~N~Livanov$^{1}$, S~M~Piyadin$^{1}$, S~G~Reznikov$^{1}$, Ya~T~Skhomenko$^{1,3}$, A~A~Terekhin$^{1}$, 
A~V~Tishevsky$^{1}$,  A~V~Averyanov$^{1}$, S~N~Bazylev$^{1}$, A~S~Belov$^4$, A~V~Butenko$^1$, E~V~Chernykh$^{1}$, Yu~N~Filatov$^5$,
V~V~Fimushkin$^1$, D~O~Krivenkov$^{1}$, A~M~Kondratenko$^6$, M~A~Kondratenko$^6$,  A~D~Kovalenko$^1$,   I~V~Slepnev$^{1}$, V~M~Slepnev$^{1}$, A~V~Shutov$^{1}$,  A~O~Sidorin$^1$,  I~E~Vnukov$^3$, V~S~Volkov$^3$}

\address{$^1$ Joint Institute for Nuclear Research, Dubna, Russia}
\address{$^2$ Physics Department, University of \v{Z}ilina, \v{Z}ilina, Slovakia}
\address{$^3$ Belgorod State National Research University, Belgorod, Russia}
\address{$^4$ Institute of Nuclear Physics, Moscow, Russia}
\address{$^5$ Moscow Institute of Physics and Technology, Moscow, Russia}
\address{$^6$ Science and Technique Laboratory "Zaryad", Novosibirsk, Russia}

\ead{$^\dagger$vladygin@jinr.ru}

\begin{abstract}
The spin program at NICA using SPD and MPD requires high intensity polarized proton beam with high value of the beam polarization.
First results on the measurements of the proton beam polarization performed
at internal target  at Nuclotron are reported.
The polarization of the  proton beam provided by new source of polarized ions 
has been measured at 500 MeV using quasielastic proton-proton scattering
and DSS setup at internal target.
The obtained value of the vertical polarization of $\sim$35\% is consistent 
with the calculations taking into account the current magnetic optics of the
Nuclotron injection line. 
\end{abstract}

\section{Introduction}
The availability of the polarized proton beam is required by the spin program at NICA \cite{spd0}. 
The main part of the physics program with SPD detector consists of the  
measurements of asymmetries in the lepton pair (Drell-Yan) production in collisions of non-polarized, longitudinally and transversally polarized protons and deuterons beams. These measurements can provide an access to all leading twist collinear and Transverse-Momentum Dependent ditsribution functions of quarks and anti-quarks in nucleons. The measurements of asymmetries in production of $J\Psi$ and direct photons, which supply complimentary information on the nucleon structure, will be performed simultaneously with Drell-Yan data using dedicated triggers. The set of these measurements permits to tests the quark-parton model of nucleons at the QCD twist-2 level with minimal systematic errors \cite{spd1,spd2}. 
This program can be extended by the measurements of the single and double spin asymmetries in the pion, kaon, proton inclusive production, 
vector mesons and hyperon production in polarized proton-proton collisions etc.
Moreover, polarized proton beam is needed for the spin studies with fixed target, namely, 
measurements of spin observables  proton-proton, proton-neutron and proton-deuteron  elastic scattering.,
investigation of the spin structure of the short-range nucleon correlations and three nucleon forces \cite{lad1}.
The  realization of this program requires good knowledge of the proton beam polarization.

In this paper we report first results of the  proton beam polarization measurements performed using
upgraded polarimeter \cite{ITS_polarimeter} at the Nuclotron internal target station. 

\section{Experiment at ITS}

The polarimeter based on the use of  $dp$- elastic scattering at large angles 
($\theta_{\rm cm}\ge 60^{\circ}$) at 270 MeV\cite{ITS_polarimeter}, where precise data on analyzing 
powers \cite{kimiko,Sekiguchi04,suda_nim} exist, has been developed at internal target 
station (ITS) at Nuclotron\cite{ITS}. The accuracy of the
determination of the deuteron beam polarization achieved with this method is 
better than 2\% because of the values of the analyzing powers were obtained for the polarized 
deuteron beam, which absolute polarization had been calibrated via the 
${\rm ^{12}C}(d,\alpha){\rm ^{10}B^*[2^+]}$ reaction\cite{suda_nim}.

The use of large amount of the scintillation counters allowed to cover
wide angular range. The measurement of the beam polarization 
has been performed at 270 MeV where the precise data on the tensor and vector analyzing
powers based on the absolute calibration of the beam polarization exist \cite{suda_nim}.
These measurements were performed using internal target station at Nuclotron \cite{ITS} with new   
control and  data  acquisition system \cite{ITS_DAQ}. 
The polarimeter has been upgraded \cite{ITS_polarimeter}
by new VME based DAQ \cite{isupov_dspin2017},
new MPod based high voltage system \cite{skhomenko, skhomenko_dspin2017}, new system of monitors etc. 

The same setup has been used to measure the vector $A_y$ and tensor $A_{yy}$ and $A_{xx}$ 
analyzing powers in $dp$- elastic scattering between 400 MeV and 1800 MeV \cite{ladygin_dspin2017} 
using polarized deuteron beam from 
new source of polarized ions (SPI) developed at LHEP-JINR \cite{NewPIS}.
The setup has been also  adopted for the measurements of the proton beam polarization at 500 MeV.

\section{Measurements of the analyzing power for $pp$- elastic scattering} 

The classical method to measure the proton beam polarization at
intermediate and high energies is the use of the left-right $pp$- elastic or quasi-elastic scattering 
(see, for instance, \cite{azhgirey} and references therein).
The maximal value of the analyzing power at the energies below 1000 MeV is close to $\sim$40$^\circ$  in cms \cite{said},
that corresponds roughly 14-15$^\circ$ in the laboratory. Unfortunately, this angle is  inaccessible due 
to design of ITS and detector support.    

This method has been modified to increase the polarimeter figure of merit. 
For this purpose the measurements of the proton beam polarization using several
pairs of the detectors placed in the kinematic coincidences corresponding to $pp$- elastic scattering in 
the horizontal plane (orbit plane of Nuclotron) have  been proposed.
The similar method has been used at COSY using EDDA setup \cite{edda1,edda2}.
The feasibility of the proposed method at ITS has been checked in November 2016 run at Nuclotron using polarized deuteron 
beam with the energy of 500~MeV/nucleon.

New SPI \cite{NewPIS} has been used to provide 
polarized deuteron beam. In the current experiment the spin modes with the maximal ideal values 
of ($P_z$,$P_{zz}$)= (0,0), (-1/3,+1) and (-1/3, +1) were used. 
The deuteron beam polarization has been measured at 270 MeV \cite{ITS_polarimeter}.
The $dp$- elastic scattering  events at 270 MeV were selected using correlation of the energy losses 
and time-of-flight difference  for deuteron and proton detectors.  
The values of the beam polarization for different spin  have been obtained  as weighted averages for
8 scattering angles for $dp$- elastic scattering in the horizontal plane only. They were measured as ($P_z$,$P_{zz}$)= 
(-0.232$\pm$0.018,+0.595$\pm$0.013) and
(-0.243$\pm$0.013,-0.736$\pm$0.011) for spin modes "2-6" and "3-5", respectively.

After deuteron beam polarization measurements at 270 MeV, the beam has been accelerated up to
1000 MeV (or 500 MeV/nucleon).
Eight pairs of the scintillation detectors were positioned in the horizontal plane 
covering angular range 55$^\circ$-125$^\circ$ in the cms
for $pp$- elastic scattering at 500 MeV on the left. Since analyzing power is antisymmetric with respect
to $\theta^*$=90$^\circ$, the scattering in the backward hemisphere on the left at the angle $\pi-\theta$ can be considered as the 
scattering in the forward hemisphere on the right at the angle $\theta$ \cite{lehar}.

\begin{figure}[h]
\begin{minipage}[t]{0.47\textwidth}
 \centering
  \resizebox{7cm}{!}{\includegraphics{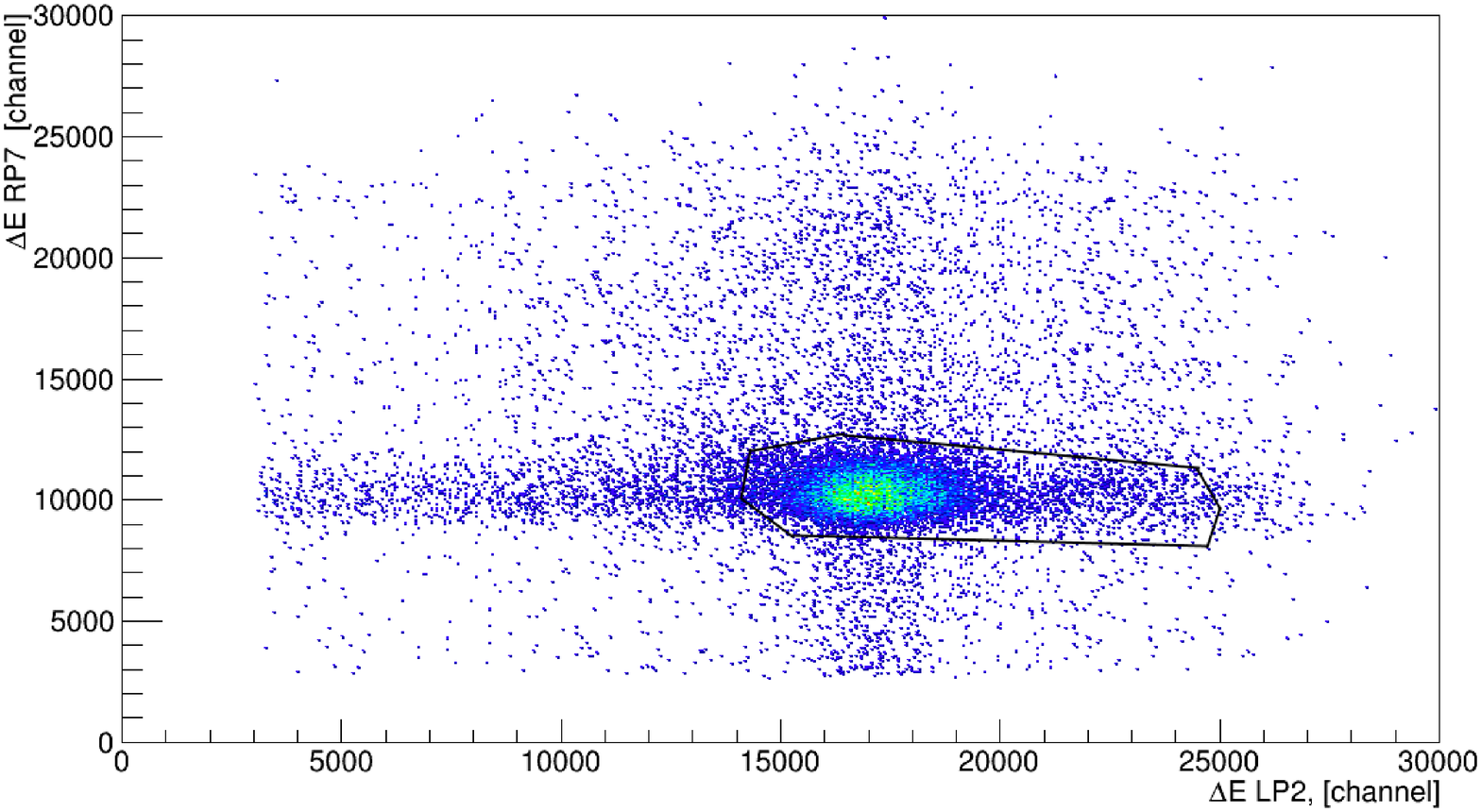}}
\caption{Correlation of the signal amplitudes in scintillation counters for proton-proton coincidences
at 65$^\circ$ in cms. The solid line is a graphical cut to select $pp$- quasielastic scattering events.}
\label{fig:fig1}
\end{minipage}\hfill
\begin{minipage}[t]{0.47\textwidth}
 \centering
  \resizebox{8.2cm}{!}{\includegraphics{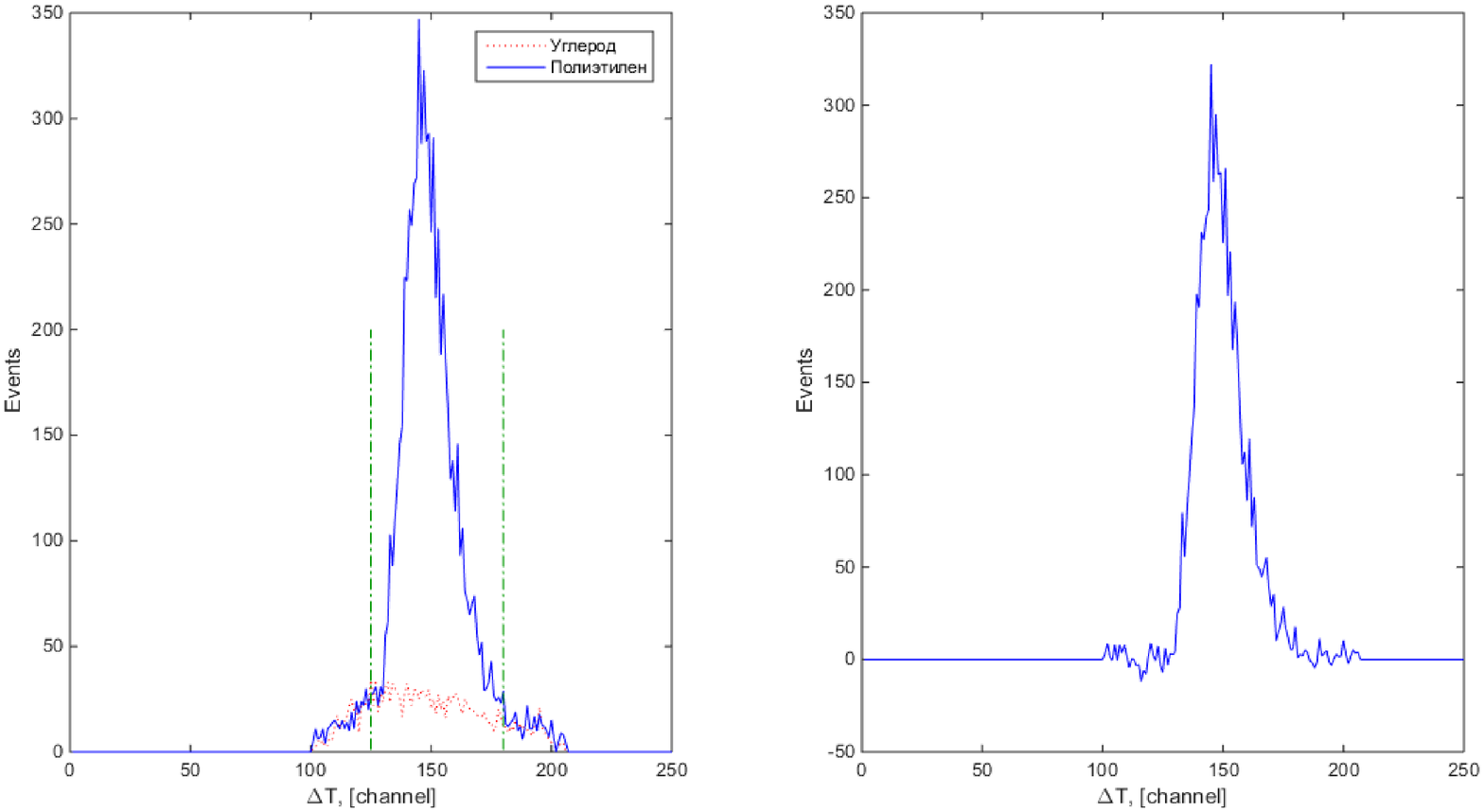}}
\caption{The time-of-flight difference  between the signals from two proton detectors placed at 65$^\circ$ in cms.
Left and right panels are the spectra obtained on CH$_2$ target and after carbon subtraction, respectively. }
\label{fig:fig2}
\end{minipage}
\end{figure}

The main part of the measurements were performed using 
CH$_2$ target. Carbon target was used  to estimate the background.
The $pp$- quasielastic scattering events were selected using the energy losses 
and time-of-flight difference for two proton detectors  placed in the kinematic coincidences. The correlation of the signal amplitudes in scintillation counters for proton-proton coincidences at 65$^\circ$ in cms is shown in Fig.\ref{fig:fig1}.
The time-of-flight difference  between the signals from two proton detectors placed at 65$^\circ$ in cms is presented in Fig.\ref{fig:fig2}.
Left and right panels are the spectra obtained on CH$_2$ target and after CH$_2$-C  subtraction, respectively.

The normalized numbers of $pp$-elastic scattering events for each spin mode were used to calculate the values of the analyzing power of     
$pp$- elastic scattering at 500 MeV. The obtained results were found to be in good agreement
with the  SP07 solution of SAID PWA \cite{said}, that proved the feasibility of the proposed method.

\section{Results of the proton beam polarization measurement}

The unpolarized and polarized proton beam provided by SPI \cite{NewPIS} has been accelerated  in March 2017 run up to 500 MeV. 
The typical intensity of the beam was $\sim$1.5$\cdot$10$^8$ ppp and  $\sim$2-3$\cdot$10$^7$ ppp for 
unpolarized and polarized cases, respectively.   
SPI provided proton beam polarization using WFT~1$\to$3 with ideal value of the polarization $P$=-1. 
The polarization of the proton beam has been obtained using the data from eight pairs of the detectors placed 
in the kinematic coincidences. The values of the 
analyzing power for $pp$ elastic scattering were taken from SAID PWA \cite{said}.  

The weighted average values of the proton beam polarization were found as 0.017$\pm$0.021 and -0.354$\pm$0.022 for
unpolarized and polarized cases, respectively. According   to spin transport calculations 
taking into account  the current magnetic option of the Nuclotron injection line \cite{filatov_dspin2017}
the polarization at the exit of SPI equals -0.90$\pm$0.06. The vertical component of the proton beam polarization 
at the ITS point can be 
increased up to this value by the installation of two solenoids at the exit of SPI and in to  the Nuclotron injection line \cite{filatov_dspin2017}.

\section{Conclusions}
The ITS deuteron beam polarimeter \cite{ITS_polarimeter} has been used to measure   the proton beam polarization at 500 MeV using  $pp$- quasi-elastic scattering.

The obtained value of the vertical proton polarization is -0.354$\pm$0.022.
This value corresponds to  polarization at the exit of SPI of -0.90$\pm$0.06 taking into account  the current magnetic option of the Nuclotron injection line. Installation of two solenoids at the exit of SPI and in to  the Nuclotron injection line will increase
the  vertical component of the proton beam polarization 
at the ITS point.  

The current version of the  ITS deuteron beam polarimeter can be applied for the proton beam  polarization measurement at the energy range of 200-1000 MeV. The extension of the proton polarimetry at ITS to the higher energies (up to 3500 MeV) is possible by the enlargement of the angular span of the polarimeter using new detector support, new scintillation counters etc. 

The availability of the polarized proton beam allows to extend the DSS physics program at ITS \cite{lad1}, namely, to perform  the experiments on the measurements of the nucleon analyzing power $A_y^p$ in  $pd$- elastic scattering   at 135-1000 MeV  and  in $pd$- nonmesonic breakup at the energies between  135-250 MeV for different kinematic configurations etc.

\ack
The authors thank the Nuclotron staff for providing good conditions of the experiment. 
They are grateful to Dr.V.A.~Mikhaylov for the advise during acceleration of polarized protons.
They thank V.B.~Shutov and V.I.~Maximenkova for the help 
during the preparation of the detector.
The work has been supported in part by the RFBR under grant $N^0$16-02-00203a and
JINR- Slovak Republic scientific cooperation program in 2016-2017.


\section*{References}
\begin{thebibliography}{99}
\bibitem{spd0} \textit{http://nica.jinr.ru}
\bibitem{spd1} Nagaitsev~A~P \textit{et al.}~{\it talk at this Conference}. 
\bibitem{spd2} Guskov~A~V \textit{et al.}~{\it talk at this Conference}. 
\bibitem{lad1} Ladygin~V~P \textit{et al.} ~2016 \textit{Int. J. Mod. Phys. Conf. Ser.} \textbf{40} 1660074. 
\bibitem{ITS_polarimeter} Kurilkin~P~K \textit{et al.} ~2011  
 \textit{Nucl.Instr.Meth. in Phys.Res.~A} \textbf{ 642}   45.

%\bibitem{uesaka} Uesaka~T \textit{et al.} ~2006 ~{\it Phys.Part.Nucl.Lett} \textbf{3} 305.
%\bibitem{gurchin1} Gurchin~Yu~V \textit{et al.} ~2007 ~{\it Phys.Part.Nucl.Lett.} \textbf{4} 263.


\bibitem{kimiko} Sekiguchi~K \textit{et al.} ~2002 {~\it Phys.Rev.} C \textbf{65} 
 034003.
\bibitem{Sekiguchi04} Sekiguchi~K \textit{et al.} ~2004 ~{\it Phys.Rev.} C \textbf{70} 014001.
\bibitem{suda_nim} Suda~K \textit{et al.} ~2007 ~{\it Nucl.Instr.Meth. in Phys.Res.} A \textbf{572} 745.

\bibitem{ITS} 
Malakhov~A~I \textit{et al.} ~2000 \textit{Nucl.Instrum.Meth. in Phys.Res.~A} \textbf{440} 320.
\bibitem{ITS_DAQ}
Isupov~A~Yu  \textit{et al.} ~2013
\textit{Nucl.Instrum.Meth. in Phys.Res.~A} \textbf{698} 127.
%Isupov~A~Yu, Krasnov~V~A, Ladygin~V~P, Piyadin~S~M and  Reznikov~S~G    ~2013
%\textit{Nucl.Instrum.Meth. in Phys.Res.~A} \textbf{698} 127.

\bibitem{isupov_dspin2017} Isupov~A~Yu ~{\it talk at this Conference}. 
\bibitem{skhomenko} Skhomenko~Ya~T \textit{et al.} ~2016~{\it Scientific Statements of Belgorod State University. Series: Mathematics and Physics} \textbf{43} 115. 
 
\bibitem{skhomenko_dspin2017} Skhomenko~Ya~T \textit{et al.} ~{\it talk at this Conference}. 
\bibitem{ladygin_dspin2017} Ladygin~V~P \textit{et al.} ~{\it talk at this Conference}. 

\bibitem{NewPIS}
Fimushkin~V~V \textit{et al.} ~2016  \textit{J.Phys.Conf.Ser. } \textbf{678} 012058.
 
\bibitem{azhgirey} Azhgirey~L~S \textit{et al.} ~2003 \textit{Nucl.Instrum.Meth. in Phys.Res.~A} \textbf{497} 340.
 
\bibitem{said} 
\textit{http://gwdac.phys.gwu.edu}
\bibitem{edda1} Altmeier~M  \textit{et al.} ~2005  \textit{Eur.Phys.J.} A \textbf{23} 351.
\bibitem{edda2} Bauer~F \textit{et al.} ~2005 ~{\it Phys.Rev.} C \textbf{71} 054002.
\bibitem{lehar} Lechanoine-LeLuc~C and Lehar~F ~1993~ \textit{Rev.Mod.Phys.} \textbf{65} 47.

\bibitem{filatov_dspin2017} Filatov~Yu~N \textit{et al.} ~{\it talk at this Conference}. 


%\bibitem{pkur2014ppn}  Kurilkin~P~K and  Ladygin~V~P ~2014 \textit{Phys.Part.Nucl.} \textbf{45}  265.


%\bibitem{dss0} Ladygin~V~P  \textit{et al.} ~2011 \textit{J.Phys.Conf.Ser.} \textbf{295} 012131.
%\bibitem{dss1} Ladygin~V~P  \textit{et al.} ~2014 \textit{Phys.Part.Nucl.} \textbf{45}  327.
%\bibitem{dss2} Ladygin~V~P  \textit{et al.} ~2014 \textit{Few Body Syst.} \textbf{55} 709.


%\bibitem{PLB2012} Kurilkin~P~K \textit{et al.} ~2012 \textit{Phys.Lett.~B} \textbf{715} 61.
%\bibitem{PPNL2011} Kurilkin~P~K \textit{et al.} ~2011 \textit{Phys.Part.Nucl.Lett.} \textbf{8}  1081.

 
\end{thebibliography}
\end{document}